\documentclass[journal]{IEEEtran}
\ifCLASSINFOpdf
\else
\fi
\usepackage{amsmath}
\usepackage{amssymb}
\usepackage{float}
\DeclareMathOperator{\sign}{sign}

\usepackage{todonotes}
\usepackage{graphicx}
\usepackage{subcaption}
\usepackage{booktabs}
\begin{document}
%


\title{Counteracting Concept Drift by Learning\\ with Future Malware Predictions}
%
%
%

\author{
\IEEEauthorblockN{Branislav Bo{\v s}ansk{\' y}\IEEEauthorrefmark{1}\IEEEauthorrefmark{2}, Lada Hospodkov{\'a}\IEEEauthorrefmark{1}, Michal Najman\IEEEauthorrefmark{1}, Maria Rigaki\IEEEauthorrefmark{2}, Elnaz Babayeva\IEEEauthorrefmark{1}, Viliam Lis{\'y}\IEEEauthorrefmark{1}\IEEEauthorrefmark{2}}\\
    \IEEEauthorblockA{\IEEEauthorrefmark{1}Avast Software / Gen
    \\\{branislav.bosansky, lada.hospodkova, michal.najman, elnaz.babayeva, viliam.lisy\}@avast.com}\\
    \IEEEauthorblockA{\IEEEauthorrefmark{2}AI Center, Dept. of Computer Science, FEE,
Czech Technical University in Prague
    \\\{branislav.bosansky, maria.rigaki, viliam.lisy\}@fel.cvut.cz }
    \thanks{The work on this paper has primarily been done in 2022.}
    }

%
%

\markboth{Counteracting Concept Drift by Learning with Future Malware Predictions, 2022}%
{Shell \MakeLowercase{\textit{et al.}}: Bare Demo of IEEEtran.cls for IEEE Journals}
%


\newcommand{\calX}{{\cal X}}
\newcommand{\calY}{{\cal Y}}
\newcommand{\calZ}{{\cal Z}}
\newcommand{\calP}{{\cal P}}


\maketitle

\begin{abstract}
The accuracy of deployed malware-detection classifiers degrades over time due to changes in data distributions and increasing discrepancies between training and testing data. 
This phenomenon is known as the \emph{concept drift}.
While the concept drift can be caused by various reasons in general, new malicious files are created by malware authors with a clear intention of avoiding detection.
The existence of the intention opens a possibility for predicting such future samples.
Including predicted samples in training data should consequently increase the accuracy of the classifiers on new testing data.

We compare two methods for predicting future samples: (1) adversarial training and (2) generative adversarial networks (GANs).
The first method explicitly seeks for adversarial examples against the classifier that are then used as a part of training data.
Similarly, GANs also generate synthetic training data.
We use GANs to learn changes in data distributions within different time periods of training data and then apply these changes to generate samples that could be in testing data.
We compare these prediction methods on two different datasets: (1) Ember public dataset and (2) the internal dataset of files incoming to Avast. 
We show that while adversarial training yields more robust classifiers, this method is not a good predictor of future malware in general.
This is in contrast with previously reported positive results in different domains (including natural language processing and spam detection). 
On the other hand, we show that GANs can be successfully used as predictors of future malware.
We specifically examine malware families that exhibit significant changes in their data distributions over time and the experimental results confirm that GAN-based predictions can significantly improve the accuracy of the classifier on new, previously unseen data. 
\end{abstract}

\begin{IEEEkeywords}
malware concept drift, adversarial training, generative-adversarial samples
\end{IEEEkeywords}

\IEEEpeerreviewmaketitle

\section{Introduction}
One of the most basic assumptions of machine learning is that training and testing data come from the same probability distribution. In many real-world problems, this is not the case and the distribution that generates the data evolves continuously. User preferences for music or movies in recommendation systems naturally change over time; the structure of web pages for categorization may quickly change with a new version of a popular tool for their creation; pictures from a stationary camera may change with weather or seasons, while we are still interested in recognizing the same objects. This phenomenon is known as ``concept drift'' (see \cite{gama2014survey} for a survey).

In security scenarios, such as email spam detection or malware detection, the natural concept drift caused by new campaigns is complemented by intentional optimization of an adversary, who tries to avoid detection by automated detection systems. Spam authors use less obvious synonyms or misspelled words to deceive the spam filters. Malware authors use obfuscation and encryption to avoid antivirus systems. Since the attackers in security scenarios can often test their attack on the current version of the classifiers, the object they create for classification in the next time intervals depends on the classifier currently deployed. This kind of concept drift was named as \emph{adversarial concept drift} \cite{Miller2014}.

The presence of concept drift has several consequences on machine learning applications: (1) the performance of a classifier trained on data available at any given time generally degrades over the following time periods; (2) too old data samples may not be relevant for training a model that should be effective in classifying future data. 
Most methods dealing with concept drift assume that changes in the data distribution happen quickly and unpredictably. In that case, the general approach is to create an ensemble of classifiers \cite{Wang2003}, training which starts at different points in time. The performance of these classifiers is then carefully monitored and the weights of their influence on the classification at any given moment depend on their performance in the recent past \cite{kolter2007dynamic}. Eventually, the worst-performing classifiers are removed and replaced by newly trained classifiers.

We conjecture that the concept drift can be, to some level, predicted in some domains. For the malware-detection domain, there are at least three mechanisms that can make concept drift predictable: (1) The first is \textbf{inertia}. If a certain malware family starts rapidly spreading in one time period, it is likely that this family will spread in the subsequent time period as well. (2) The second is \textbf{adoption of new features} used by other malware authors. When malware starts exploiting a new vulnerability to infect computers, a new monetization vector, or a new type of encryption, it is likely that other malware authors will learn about it and will start using the same or similar methods. (3) Third reason why concept drift may be predictable is \textbf{adversarial adaptation}. If malware is detected by major antivirus systems, its spread will be severely limited and the revenue for the authors will decrease. Due to computational reasons, many anti-virus systems often rely on detecting simple patterns in the file or its behavior. 
Hence, the authors are strongly motivated to make small inexpensive changes to the malware that will disrupt the patterns used for detection. 

Our main contribution is the evaluation of concept-drift prediction in the malware-detection domain.
First, we define the appropriate evaluation methodology: (1) we split the data in time -- the training window for the classifier corresponds to given time periods and the testing data are from subsequent time periods; (2) for each such time split, we compare the performance of models trained in a standard way with models that use some predictions of concept drift,  thus simulating realistic setting with regularly retrained models.
We use two methods for predicting samples: (1) adversarial samples found as small modifications of features of malicious samples such that they evade the classifier (known as \emph{adversarial training}) and (2) samples generated by generative adversarial networks (GANs) that are trained to mimic changes observed in samples over time in the past. 
We compare the methods on two datasets: (1) public Ember dataset~\cite{anderson2018ember} and (2) an internal dataset created from files incoming into Avast. Our experimental results show that including GAN-predicted samples into training data can significantly improve the accuracy of the classifiers on future unseen data. 

\section{Related Work}
The negative effects of the concept drift are well known. In the context of \emph{adversarial concept drift}, several works study the possibilities of generating and using adversarial samples during the training to counteract the concept drift. The main differences compared to existing works are: (1) we adapt the methods for generating adversarial samples to the malware-detection domain, (2) we provide a methodology for measuring the realistic impact of methods counteracting the concept drift by comparing to regularly retrained model as a realistic baseline, (3) we show that our GAN models are able to improve the accuracy on both datasets, the public one (Ember) where the improvement is minor, and a more realistic internal dataset, where the improvement was up to $10\%$ in true positive rate in some cases. 

\subsection{Adversarial Training}
Finding adversarial samples that evade the current classifier and adding them into the training data is known as \emph{adversarial training}~\cite{szegedy2014intriguing,deepfool, nguyen2014, jordan}.
Many approaches for generating adversarial samples modify the genuine sample by adding a small perturbation which moves the sample towards the desired class. A suitable direction of the movement is estimated with the gradient of a training loss with respect to the genuine sample. 
It has been shown that the use of adversarial training spans beyond the traditional domain of images to natural language processing \cite{samanta2017crafting, sun2020advbert} as well as to malware-detection domain~\cite{Grosse2017}.

The idea of using adversarial training to counteract concept drift has been explored in other domains. 
In email-spam detection~\cite{bruckner2012static}, the authors propose game-theoretic (adversarial) training of Logistic Regression (LR) and Support Vector Machine (SVM) that takes into account the prediction of future actions by the attackers.
The empirical evaluation shows that the resulting classifier is much more robust against concept drift.
While the F-measure of the classifiers trained in the standard way degraded up to 17\% over time, the adversary-aware classifiers degraded only up to 7\%. A similar study is presented in \cite{colbaugh2012predictive}, where they compare Naive Bayes (NB) classifier with a linear classifier trained by game-theoretic min-max optimization. While the accuracy of the NB classifier degraded from 97\% to 75\% over four years, the accuracy of the game-theoretic classifier stayed over 95\% throughout the whole testing period. Our goal in this paper is to verify whether the methods of adversarial training achieve similar results in the malware-detection domain and compare them to a regularly retrained classifier.


For many domains, including malware detection, there are two main approaches for generating adversarial samples: (1) to generate the samples in the feature-space only or (2) to generate adversarial samples that correspond to an actual object (e.g., a malicious Windows portable executable (PE)). 
The problem-space adversarial attacks can be more realistic~\cite{pierazzi2020}, however, generating problem-space attacks is either much more difficult or these attacks are restricted only to a specific type (e.g., only using carefully selected modifications of Windows PEs that surely do not alter the behavior of the PE~(e.g., in \cite{castro2019armed})).
Therefore, we focus on adversarial training with feature-based attacks since for rich and well-chosen feature space it was shown that feature-space attacks can still result in classifiers robust also against problem-space attacks~\cite{tong2019improving}.

\subsection{Generative Adversarial Networks (GANs)}
Similar to other domains, GANs in the malware-detection domain are used for generating adversarial samples \cite{Li_GAN_android} or to augment data sets. 

For counteracting the concept drift, GANs have recently been used in \cite{dai2021anticoncept}. The authors show that the degradation of the model that uses artificial adversarial samples generated by GANs is significantly smaller compared to the baseline training (the accuracy is higher by $10\%$ in more distant future time periods).
However, the experimental evaluation is not performed to such details to confirm that the proposed predictor would outperform the regularly retrained classifier in practice. 
Instead, we (1) give a novel and detailed methodology for training and using GANs as predictors that is not described in the work above, and (2) our experiments clearly show that including the predictions of GANs into training data can improve the accuracy of the regularly retrained model.

Other researchers used generative models to augment the dataset but not with the goal of counteracting the concept drift. 
Burks et al.~\cite{Burks2019} use GANs and Variational Autoencoder (VAE) to generate additional samples of underrepresented families and report an increase of accuracy on testing data by 7\%. While our approach shares some ideas (we are using GANs to augment data sets), we are explicitly training GANs to learn changes (transformations) in the data distribution of a particular malware family in the past and then apply these transformations to training data to get predictions of the testing data distribution. Moreover, we are explicitly testing our models on datasets that allow for the time-split evaluation -- we use several time periods as training data and the next time period as testing data.

Kim et al.~\cite{Kim2018} use image representation of binaries and GANs to improve the detection of zero-day attacks. To simulate the zero-day attacks, the authors add noise into the image, however, they are not using new, unseen versions of the malware files to verify whether their models can counteract the effects of concept drift.

\subsection{Domain and Out-Of-Distribution Generalization}
The problem of generalization to new unseen data is also deeply studied outside the security domain and there are many machine-learning communities focused on this challenge. \emph{Domain generalization}~\cite{zhou2022domain,wang2022generalizing} often focuses on creating a model that can perform well on unseen, novel data distributions without requiring explicit knowledge or access to those distributions during training. Methods studied in domain generalization typically involves training on multiple diverse domains, employing strategies to extract and generalize the underlying features that are invariant across these domains. \emph{Out-of-distribution generalization}~\cite{liu2021towards,Hendrycks_2021_ICCV,yang2021generalized}, on the other hand, extends the concept of domain generalization by focusing on the model's ability to perform well on any data that comes from a distribution different from the training distribution, regardless of whether these are categorized into specific "domains" or not. The challenge here is not just generalizing across several known domains but ensuring robustness against all possible shifts in the data distribution.

Our methods generate new samples that are incorporated into our training data, hence they can be viewed as a form of data augmentation, similar to well-known techniques used in domain generalization or out-of-distribution generalization (e.g., Mixup~\cite{zhang2017mixup}, CutMix~\cite{yun2019cutmix}, or style adaptation~\cite{somavarapu2020frustratingly}). However, in our context, we create new samples specifically to improve the model's ability to handle concept drift in the data (either Ember or internal Avast data). Additionally, our objective is to \emph{predict} the adversarial concept drift and assess whether strategies such as adversarial training or modeling past data shifts with GANs can effectively mitigate this drift.

\section{Problem Statement and Methodology}\label{sec:problem_statement}

Studying concept drift and evaluating methods that mitigate the effects of concept drift requires that samples from the dataset have assigned timestamps. 
Such a dataset can be split into the training part and the testing part, where the testing part follows in time after the training part. 
Moreover, it is necessary to have enough data spread over a longer time period so that multiple such splits can be done.
Having such multiple cases of training/testing data simulates comparison to a baseline approach -- regularly retraining the classifier (termed \textsc{Normal}).
Any successful method for mitigating the effects of concept drift should outperform such a baseline.

Formally, let $\calX$ be a set of samples and let $\{T_1, T_2, \ldots, T_N\}$ be the partitioning of the samples into $N$ time periods, such that $T_i \subset \calX$, $T_i \cap T_j = \emptyset$ for all $i,j \in \{1, \ldots, N\}, i \neq j$, and $\bigcup_i T_i = \calX$. 
Moreover, samples $T_i$ from time period $i$ have earlier timestamps than any sample from time periods $j \in \{i+1, \ldots, N \}$.
Let's assume that for each time period~$i$, the samples are drawn from a potentially different distribution~$\calP_i$. For some sizes of training and testing window ($w_1$ and $w_2$, respectively), and time period $k \in \{1, \ldots, N\}, w_1 < k$, samples from partitions $T_{k-w_1}, T_{k-w_1+1}, \ldots, T_{k-1}$ form the training set and $T_{k}, \ldots, T_{k+w_2}$ form the testing set. A classifier is a function $\varphi_\theta: \calX \rightarrow \calY$, where $\calY$ is a set of labels. For simplicity, we assume that $\calY = \{\textit{benign, malware}\}$. Finally, we use a threshold for the decision of the classifier in order to control the false-positive rate of the classifier. This follows a common practice in cybersecurity applications where false positives can be extremely costly~\cite{kshetri2021economics}. 

To evaluate the methods, we are going to train classifiers for multiple different time periods $k$ in order to simulate repeated retraining of the model and evaluating on new, unseen data from testing distributions. 
We compare the classifiers on the same testing data and focus on standard metrics such as accuracy, F1 score, and the true-positive rate of a classifier with a threshold ensuring a specific false-positive rate.
However, given the concept drift present in data, it can happen that the testing data distributions for some testing time periods are simply fundamentally more difficult to learn compared to the previous time periods. To be able to evaluate how good the proposed methods of counteracting concept drift are, we need to know what would be the performance of a classifier that would use the most accurate predictions of the concept drift. 
To this end, we introduce the second baseline, termed \textsc{UpperBound}, that also uses a subset of samples from testing time periods, i.e. $T_{k}, \ldots, T_{k+w_2}$, as training data. 
In general, the difference between the performance of the classifier that is regularly retrained (\textsc{Normal}; corresponds to the lower bound) and the performance of the classifier with optimal predictor (\textsc{UpperBound}; the upper bound) demonstrates possible impact \emph{any} method counteracting concept drift can have.

\section{Counteracting Concept Drift}\label{sec:method}

We now describe the methods for predicting new data to be included into the training process and thus counteract the negative effects of concept drift: (1) adversarial examples generated during adversarial training, and (2) samples generated by generative adversarial networks. 

\subsection{Adversarial Training}
In general, a classifier tested on adversarial examples achieves lower classification performance than on genuine data~\cite{wong2020fast}. The goal of adversarial training is to take into account the built-in adversarial susceptibility of neural networks trained on genuine data and to prevent it by injecting adversarial samples into training data. This method of augmenting the training data yields classifiers more robust against adversarial samples at the cost of genuine data test accuracy~\cite{madry2019deep}. 

A part of the concept drift in the malware-detection domain is caused by adversarial attacks aiming at avoiding the detection. Since adversarial samples are motivated by the identical principle, the adversarial training may harden the classifier against the concept drift. However, since the goal of the adversarial methods for generating adversarial samples is primarily to find an adversarial sample against some classifier, there are no guarantees that the crafted samples align with future malware.
Consequently, adversarial training may harden the classifier against hypothetical samples that are never observed in future data.

We use the most common adversarial methods: the Fast Gradient Sign Method (FGSM), the Projected Gradient Descent (PGD), and their multi-step versions to generate adversarial samples, here referred to as $k$-FGSM (often called Basic Iterative method, BIM \cite{wong2020fast}) and $k$-PGD. Our multi-step versions of adversarial methods simply repeat the adversarial step $k$-times.

In the equations below, $\ell$ is a loss that a classifier with weights $\theta$ receives on a genuine sample $x \in \calX$ belonging to a class $y \in \calY$.
The input samples are normalized such that there is zero mean value and a unit variance. 

\subsubsection{FGSM}
FGSM uses the sign of the gradient to modify the sample $x$ within an $\epsilon$-ball. We use $\epsilon \in \{0.1, 1.0\}$ depending on the dataset. The adversarial samples are found as follows: 
\begin{equation}
    x_{adv} = \Pi(x + \epsilon \cdot \sign(\nabla_x\ell(x, y, \theta))),
\end{equation}
where $\Pi : \calX \rightarrow \calX$ is a projection of the adversarial sample to an admissible space. 
Note that we apply this projection also in the case of FGSM, although this method is commonly used without such projections in the literature \cite{goodfellow2015explaining, wong2020fast}. The main reason is that the lack of this projection often produced adversarial samples that were irrelevant to the space of admissible samples and thus could have only a limited effect on counteracting the concept drift (e.g., adversarial samples could have corresponding unnormalized feature vector containing negative values that semantically correspond to values from a histogram). 

\subsubsection{PGD}
PGD modifies the genuine sample with a gradient scaled by a scaling constant $\epsilon$; we use $\epsilon = 1.0$. The adversarial samples are found as follows:

\begin{equation}
    x_{adv} = \Pi(x + \epsilon \cdot \nabla_x\ell(x, y, \theta))
\end{equation}

\subsection{Generative Adversarial Networks (GANs)}
GANs \cite{goodfellow2014generative} consist of two networks, a generator $G: \calZ \rightarrow \calX$ and a discriminator $D: \calX \rightarrow \mathbb{R}$. The generator $G$ transforms a source domain $\calZ$ to a target domain $\calX$ while the discriminator tells apart whether a vector $x \in \calX$ comes from a reference genuine data distribution $\calP_i$ or whether it is crafted by the generator. For instance, generators may transform random noise into images, effectively allowing parametric image generation \cite{karras2018progressive}, while the discriminator judges the distinctiveness of generated images from genuine data \cite{Creswell_2018}. GANs are defined as a min-max problem as in Eq. \eqref{eq:ganeq} with the criterion $L(G, D)$ as in Eq. \eqref{eq:ganloss}. The loss $L(G, D)$ incentivizes the generator $G$ to shape its output to match the reference genuine data distribution \cite{goodfellow2014generative}.

\begin{equation}\label{eq:ganeq}
    \min_G \max_D L(G, D)
\end{equation}
\begin{equation}\label{eq:ganloss}
    L(G, D) = \mathop{\mathbb{E}}_{x \sim \calP_i} \log(D(x)) + \mathop{\mathbb{E}}_z \log(1 - D(G(z)))
\end{equation}

GANs can be used for generating artificial malware samples that are (in the feature representation) indistinguishable from the actual malware.
In order to counteract the concept drift, the goal of GANs is to mimic samples from testing distributions. 
Specifically, for a time period $k$, the GANs should generate samples that are similar to testing time periods $T_k,\ldots,T_{k+w_2}$ based on the training data from training time periods. 
This way, the generative networks act as predictors of the concept drift -- given input training data from distributions $\calP_{k-w_1},\ldots,\calP_{k-1}$, GANs are generating samples similar to the testing distributions $\calP_{k},\ldots,\calP_{k+w_2}$.
Therefore, we train a collection of generative networks for multiple such time periods $k$ on our datasets and create multiple predictors of the concept drift in data. 

However, the changes in the complete data distributions $\calP_i$ can be too complex to be modeled by a simple predictor. 
The main reason is that malware files fall into different categories (e.g., there are file infectors, ransomware, etc.) and there are different malware families with different update cycles and/or lifespan. 
Therefore, we condition generative concept drift predictors with malware family labels to learn and predict the changes occurring separately in each malware family. 
Moreover, it is common for the malware authors to adopt successful routines from other families~\cite{Calleja19}. 
Hence, changes learned for malware family 1 can be also applied to training data of malware family 2 simulating the adoption of a feature/technique.

To use generative networks in this way, we need to solve several technical challenges. GAN training, as put in Eq. \eqref{eq:ganeq}, is malformed because it allows the generator to always generate the same output from the distribution $\calP_i$, completely circumventing the desired generative capability \cite{Zhu2017}. 
This situation can occur in our case where the input of the generative model is a sample from the training distribution and the goal of GANs is to generate its modification under the concept drift.
CycleGANs \cite{Zhu2017} learn to transform between two unpaired reference sets of genuine examples by introducing a backward discriminator and a backward generator that operate in the opposite direction than their standard GAN counterparts. That is, the backward generator $G_b: \calX \rightarrow \calZ$ transforms the domain $\calX$ to the domain $\calZ$ and the backward discriminator $D_b: \calZ \rightarrow \mathbb{R}$ tells apart generated vectors $z \in \calZ$ from source domain genuine examples. To enforce a vector $z \in \calZ$ is reconstructed with chained generators $G_b(G(z))$ and $x \in \calX$ is reconstructed with $G(G_b(x))$, the main CycleGAN loss $L_{cyc}$ in Eq. \eqref{eq:cyclegan} imposes an $l_1$ norm on the reconstruction in both directions. The hyper-parameter $\lambda$ indicates the importance of cycle consistency. In our experiments, we use $\lambda = 1.0$. 

\begin{equation}\label{eq:cyclegan}
\begin{split}
    L_{cyc}(G, D, G_b, D_b) & = L(G, D) \\
    & + L(G_b, D_b) \\
    & + \lambda \cdot (\,  \mathbb{E}_x \lVert x - G(G_b(x)) \lVert_1\\
    & \qquad + \mathbb{E}_z \lVert z - G_b(G(z)) \lVert_1 \,) 
\end{split}
\end{equation}
A trained CycleGAN is a solution to a min max problem in Eq. \eqref{eq:cycleganeq}.
\begin{equation}\label{eq:cycleganeq}
    \min_{G, G_b} \max_{D, D_b} L_{cyc}(G, D, G_b, D_b)
\end{equation}

\begin{figure}[t]
    \centering
    \includegraphics[width=0.48\textwidth]{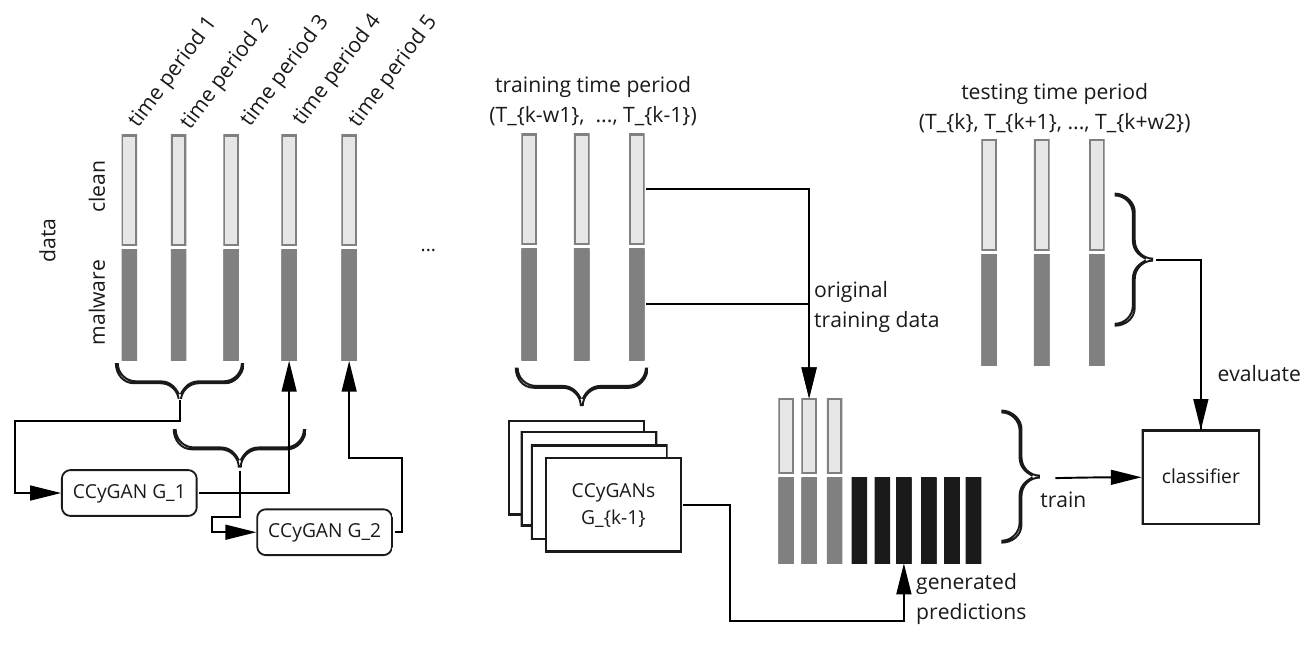}
    \caption{Schema of our approach: Conditional Cycle GANs (CCyGANs) $G_1,\ldots,G_{k-1}$ are trained to predict malware samples from subsequent time periods and then used to generated predictions from training data. The predictions are then combined with original training data. Testing data are from the time periods following the training window.}
    \label{fig:schema_gans}
\end{figure}

As discussed, malware samples $\calX$ can be additionally labeled according to malware family and this label can be used by GANs to learn concept drift occurring within a single malware family.
Conditional CycleGANs \cite{lu2018attributeguided} exploit such finer structure of the problem and modify all networks to receive a category $y' \in \calY'$ on the input, i.e. the discriminator becomes $D(x, y')$ and the generator becomes $G(z, y')$. Analogously for the backward counterparts \cite{Mirza2014}. 
We use such conditional CycleGANs (CCyGANs) for training predictors of the concept drift. For 
each time period $k'$, there is one such generative predictor $G_{k'}$ trained from previous time periods (see Fig.~\ref{fig:schema_gans} for overall schema; we use data from $\{T_{k'-w_1},\ldots,T_{k'-1}\}$ for training $G_{k'}$).
Afterward, trained CCyGANs are used to generate predicted data.
For a given time period $k$ indicating the time split, let's assume that $X_k \subset \bigcup_{i=k-w_1}^{k-1} T_i$ is the set of all samples from the training time periods, $X_k^m \subset X_k$ is the subset of these samples with \emph{malware} label, and $\{G_{i=1},\ldots,G_{i=k-1}\}$ be the set of all GANs trained to predict the concept drift in the training data. 
GANs are then used to generate additional training data $X_k^g = \lbrace G_i(x, y') | \forall x\in X_k^m, \forall y' \in \calY', \forall i \in \{1,\ldots,k-1\}\rbrace$.
Finally, we train the model in a standard way with training data $X_k \cup X_k^g$ and evaluate the model on testing time periods $T_k,\ldots,T_{k+w_2}$.

\section{Experiment Settings}\label{sec:experiments}
In this section, we describe the settings of the experiments.
First, we describe the two datasets we use for evaluating the described methods and their effectiveness against concept drift.
We use static features of Windows portable executables (PE) as defined in the Ember malware dataset~\cite{anderson2018ember}.
We first describe the common settings for both datasets following with specifics.
In all settings, we train the models repeatedly on different training time periods while the testing data are immediately following the training periods. 

We use Ember features of Windows PEs that cover a wide range of information, such as bytes and entropy histograms, strings, PE header, imports, exports, etc. 
The full length of each feature vector is $2381$. 
Due to the high and uneven variance of different features, we always normalize the data such that we clamp the extreme values to percentiles $0.01$, respectively $0.99$, and then standardize the values to have unit variance and zero mean.

Finally, we randomly split the data in each time-period $T_i$ into training, validation, and testing part -- $70\%$, $20\%$, and $10\%$.
Training data are used for training the models and the validation data are used for model and threshold selection. Finally, the testing data (from future time periods) are used for testing.
The main reason for this additional division of data within time periods is due to the \textsc{UpperBound} training -- when training an \textsc{UpperBound} model for training time-periods $\{k-w_1, \ldots, k-1\}$, data from the training part of time periods $\{k, \ldots, k+w_2 \}$ are included into training data. The model is evaluated on the testing part of $\{T_{k}, \ldots, T_{k+w_2} \}$.

\subsection{Ember 2018}
\textsc{Ember2018} is the second version of the ember malware dataset \cite{anderson2018ember}. It contains the extracted features of one million Windows PE files accessed before or during 2018. 
Each sample contains a timestamp -- a month in a year -- hence also time-periods $T_k$ for \textsc{Ember2018} correspond to months.
The dataset itself is already split between training and testing sets, however, we are using the dataset as a whole and using the training-testing split as described above.

In addition to labels such as \textit{malicious}, \textit{benign} or \textit{unlabeled}, \textsc{Ember2018} also contains malware family designations that were generated using \textit{avclass}~\cite{sebastian2016avclass}. 

To train models on \textsc{Ember2018}, we use data from 3 time-periods (months; i.e., $w_1=3$).
Testing is then performed on the subsequent single time period (i.e., $w_2=0$).
We thus create 9 separate cases where the training periods are from Jan-Mar to Sep-Nov.

\subsection{Avast Internal Dataset -- \textsc{Internal2019}}
\textsc{Ember2018} is not a dataset built specifically for studying the concept drift.
There are two main issues: (1) timestamps are very coarse (one month is a very long time period in the malware detection domain) and thus very few data points, (2) data samples are not evenly spread over the time periods, thus causing an imbalance when we consider only a subset of the time periods.

Therefore, we have created a private Avast internal dataset of $3,679,159$ Windows PE files with an exact timestamp of the observation. 
Samples were collected from Avast antivirus systems and span a period of six months. All PE files have been seen by the internal Avast framework from July 2019 to Dec 2019. For each sample, there is an assigned label malware/clean. For malware samples, we also extract the malware family designation.

A large number of all incoming files into Avast systems prevent us from effectively using all of these files in the dataset as this would be too large. 
To support the analysis of concept drift, we have created the dataset by subsampling approximately $10,000$ clean and $10,000$ malware samples for each day from July 1, 2019 to Dec 31, 2019. 
For each file, we have extracted Ember features by the LIEF library version 0.9.0~\cite{LIEF}. 

Finally, we use 7 days (1 week) as the single time period for the purpose of concept-drift experiments for the \textsc{Internal2019} dataset.
For training, we use data from 4 time-periods (4 weeks; $w_1 = 4$).
The testing is performed again on the subsequent single time period ($w_2 = 0$).
Overall, \textsc{Internal2019} offers significantly more training cases with more balanced data compared to the \textsc{Ember2018} dataset.

\begin{figure}
            \centering
            \includegraphics[width=0.49\textwidth]{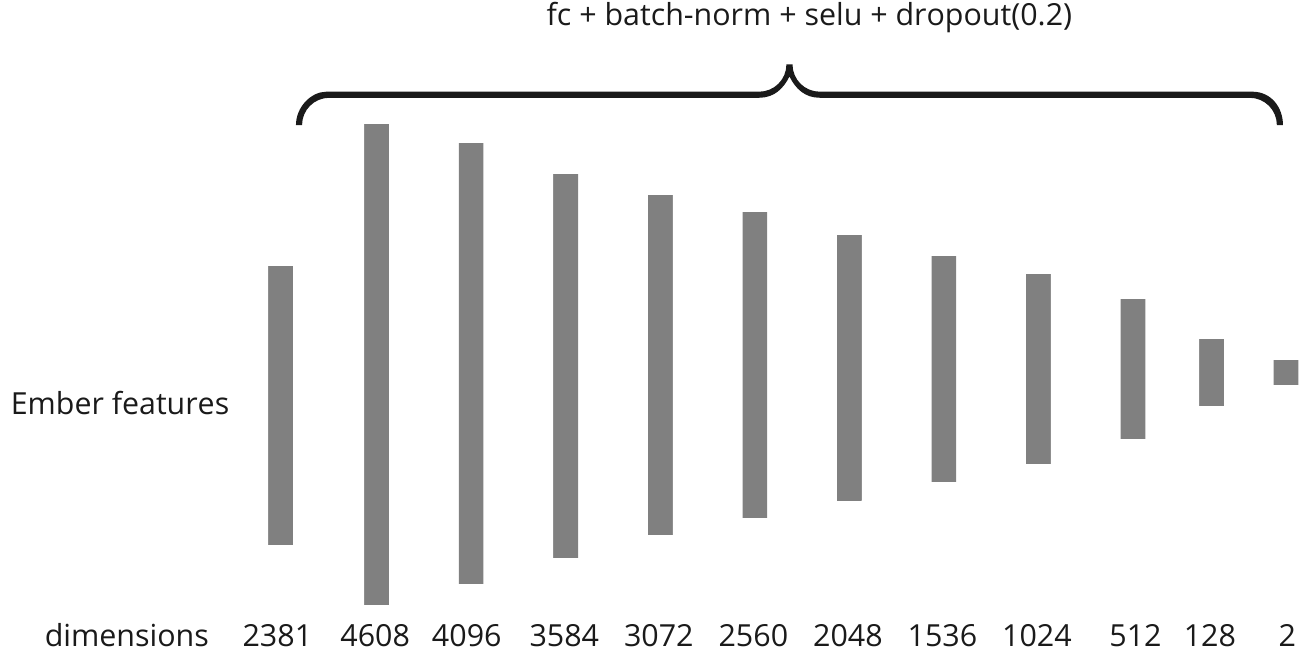}
            \caption{Architecture of neural networks used for experiments with full feature set. All layers are fully connected (denoted~\texttt{fc}), with batch normalization, selu activation function, and dropout regularization (with parameter set to $0.2$). }
            \label{fig:full-nn}
\end{figure}

\section{Experimental Evaluation}
First, we evaluate adversarial training and evaluate whether models trained using adversarial training can mitigate the effects of concept drift in the malware detection domain.
We analyze the presence of concept drift in our datasets, show the model degradation over multiple time periods, and finally compare normal training and two adversarial training methods in their ability to predict future data. 

    \begin{figure*}[t!]
        \centering
        \begin{subfigure}[b]{0.48\textwidth}
         \centering
          \includegraphics[width=\textwidth]{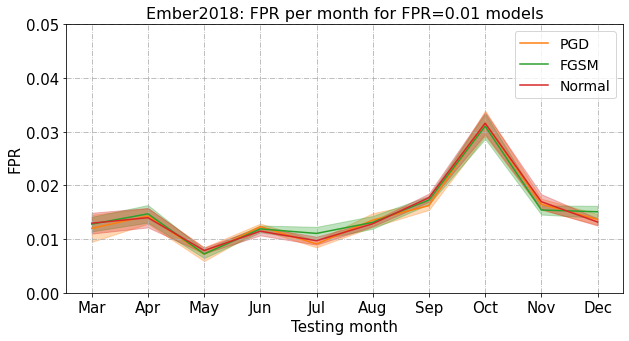}
         \caption{\textsc{Ember2018}}
         \label{fig:ember2018_fpr}
    \end{subfigure}
    \begin{subfigure}[b]{0.48\textwidth}
         \centering
          \includegraphics[width=\textwidth]{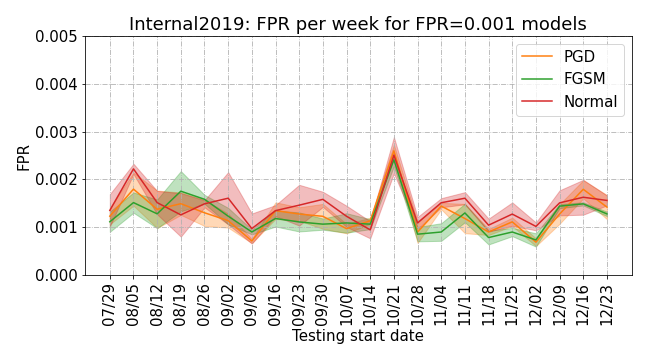}
         \caption{\textsc{Internal2019} dataset}
         \label{fig:internal_fpr}
     \end{subfigure}
     \caption{FPR per testing period using natural samples (normal model) and adversarial samples (FGSM and PGD models) for \textsc{Ember2018} (left subfigure) and \textsc{Internal2019} dataset. The shaded regions visualize the standard error of the mean.}
    \label{fig:FPR}
    \end{figure*}
    
Then we turn to the second part of the experiments -- determining whether generative models can provide useful predictions that mitigate the effects of concept drift.  
The generative models, however, are known to be computationally expensive and challenging to train properly.
Our goal is to determine whether these models can help in counteracting the concept drift, hence we restrict to only 100 features that were selected using Lasso feature selection method~\cite{Lasso2014}.
Moreover, we restrict the experiments in the second part only to families that exhibit significant changes in data distributions $\calP_i$ across different time periods.
~\\~\\
\noindent\textbf{Does Adversarial Training Counteract Concept Drift?}\\
For the experiments with adversarial training, we use the complete feature set of $2381$ features. We use a feed-forward neural network with $10$ fully connected hidden layers~(see~Fig.~\ref{fig:full-nn}). 
After each layer, we use batch normalization, \emph{selu} activation function, and \emph{dropout} (with rate~0.2). 

We use the Adam optimizer (0.01 initial learning rate) and cross-entropy loss function.
The sizes of minibatches differ depending on the datasets (the sizes were selected based on hyperparameter optimization) -- we use $128$ samples in a minibatch for \textsc{Ember2018} and $512$ samples for \textsc{Internal2019} dataset. Minibatches are always balanced -- i.e., 50\% of samples in a minibatch are malware samples, 50\% are clean samples. 
In the case of adversarial training, we only generate adversarial samples from malware samples.
Our goal is to make the training process to be as similar as possible for both normal and adversarial training.
Hence, assume that for a minibatch in normal training, there are $N$ clean samples and $N$ malware samples. 
For adversarial training, the minibatch is created such that there are $N$ clean samples listed twice, then there are $N$ malware samples, and $N$ adversarial samples created from the malware samples.

We let the training continue for at most 50 epochs (the training loss function typically leveled up well before the end of this limit) and selected the best models according to the performance on validation data. 
We select a model that maximizes the true positive rate (TPR) for a threshold set to reach a fixed FPR on validation data.
For all settings, we use 3 different FPR values -- $\{0.1, 0.01, 0.001\}$ -- i.e., we store 3 different models for a single training setting. 
To improve the clarity, we present figures corresponding only to a single threshold for each case (dataset and experiment) in the main text. 
The thresholds are selected to highlight important differences -- typically, we present the results for the threshold corresponding to FPR at $0.001$ for  \textsc{Internal2019} dataset and FPR at $0.01$ for \textsc{Ember2018}. 
A complete list of figures corresponding to all FPRs settings is listed in the supplementary materials. 
Finally, every setting was repeated at least 3 times using different seeds for the initialization of the model as well as for the data splits into the train, validation, and test sets.

Before we compare the quality of the models on the testing data, we need to verify that the FPR on testing data follows the selection made on the validation data -- i.e., that selecting a threshold for classifiers according to FPRs on the validation data results in comparable FPRs on the testing data regardless of the training method. 
The main reason for this verification is that a method with a higher FPR can also have a higher TPR. Therefore, if we want to compare TPRs on testing data, we need to confirm that the deviations among different methods are insignificant. 

\subsection{False Positive Rates on Testing Data}
\begin{figure*}[t!]
    \centering
        \begin{subfigure}[b]{0.48\textwidth}
         \centering
          \includegraphics[width=\textwidth]{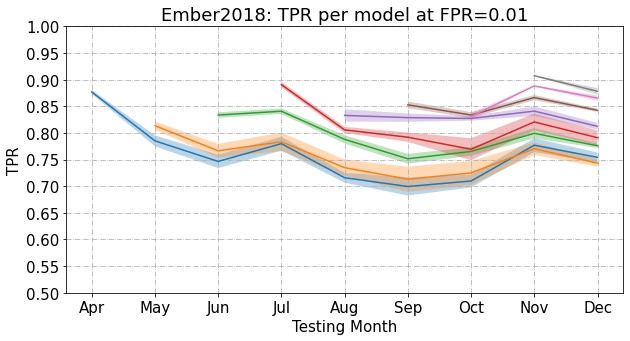}
         \caption{\textsc{Ember2018}}
         \label{fig:ember2018_degradation}
    \end{subfigure}
    \begin{subfigure}[b]{0.48\textwidth}
    \centering
    \includegraphics[width=\textwidth]{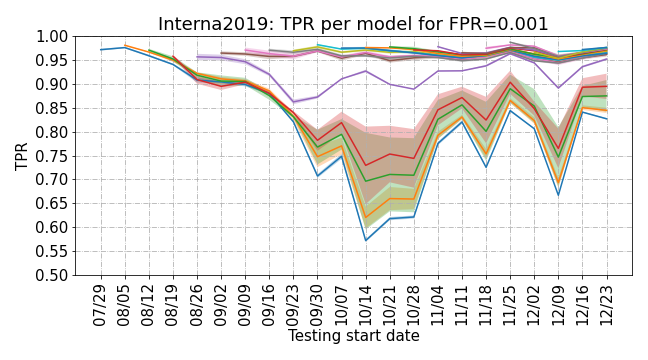}
    \caption{\textsc{Internal2019}}
    \label{fig:internal_degradation}
    \end{subfigure}
    \caption{Model degradation over multiple time periods for \textsc{Ember2018} and the \textsc{Internal2019} dataset. Each model is trained for a time period ending at the designated month/week and the TPRs for all subsequent time periods are visualized. The shaded regions correspond to the standard error of the mean.}
    \label{fig:degradation}
\end{figure*}

We compare the actual FPRs on the testing data when models use thresholds based on an FPR set according to the validation data. 

As indicated above, we use the validation data to select a binary decision threshold according to a fixed FPR level for each training time period. Fig.~\ref{fig:FPR} shows the results for the models using both the normal training as well as the adversarial training (using the two previously described methods -- FGSM and PGD). 
The measured FPRs are very similar to each other regardless of the chosen training method -- the actual values of the FPRs oscillate slightly above the target value ($0.01$ and $0.001$, respectively) with a few exceptions (such as the testing month Oct for \textsc{Ember2018} and the testing week starting Oct 21 for \textsc{Internal2019}).
In these cases, however, all models exhibited significantly higher FPR and there was no difference based on the training method.
Similar overall results hold for other FPR-based thresholds as well with two exceptions.
For \textsc{Internal2019} and FPR at 0.1, FGSM reached a significantly higher FPR (0.22) for one testing week in Dec (starting Dec 16). 
Besides these occasional exceptions, the reported FPR levels on testing data are in general comparable across different methods and thus we conclude that we can safely use and compare the quality of the models based on the reported TPRs according to a threshold that maintains a certain FPR level. 

\subsection{Model Degradation}
Next, we investigate the degradation of models trained in a standard way over longer time periods by using several testing timesteps.
This experiment demonstrates the necessity of regular retraining and deployment of new models.
Similarly, it shows the negative impact of neglecting the regular retraining can have on malware datasets.
Fig.~\ref{fig:degradation} shows the TPRs for different testing periods. 
Different lines correspond to models trained on different training windows -- therefore, each line starts in the testing period following immediately after the training window time period.

The results show there is a significant drop in the TPRs of older models for both datasets.
Consider, for example, the last 4 testing months for \textsc{Ember2018} where the models are roughly sorted based on the end of the training period.
The oldest model (blue line) is typically the worst one, and the difference in its quality to the most recent models is often as high as 15\% TPR. 
The differences are even more apparent on the \textsc{Internal2019} dataset.
The quality of the oldest models (trained in the time periods from July and the beginning of August) degrades quite quickly and their TPR drops to $0.60 - 0.75$ on testing weeks in October, while more recent models (trained on data from September) have TPRs above $0.95$.
Interestingly, the quality of the models did not degrade monotonically in  \textsc{Internal2019} and even old models reached TPR above $0.8$ for the testing week at the end of November.

The results support a well-known fact that the deployed models should be regularly retrained and redeployed. The main conclusion, however, is the observation that any evaluation of methods for counteracting the concept drift needs to take this retraining into consideration and thus we compare the models only on a single testing time period assuming that all models are going to be retrained for the next testing period.

\subsection{Adversarial Training vs. Normal Training}
\begin{figure*}
            \centering
            \includegraphics[width=0.48\textwidth]{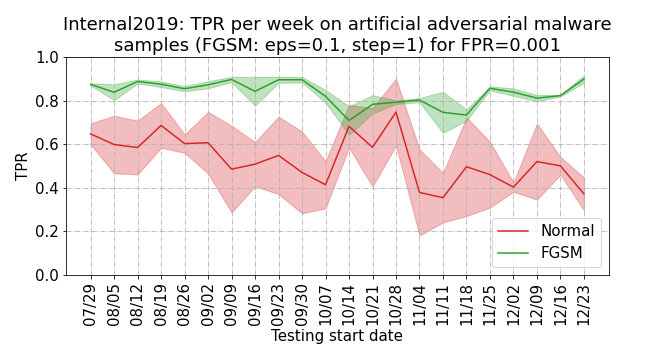}
            \includegraphics[width=0.48\textwidth]{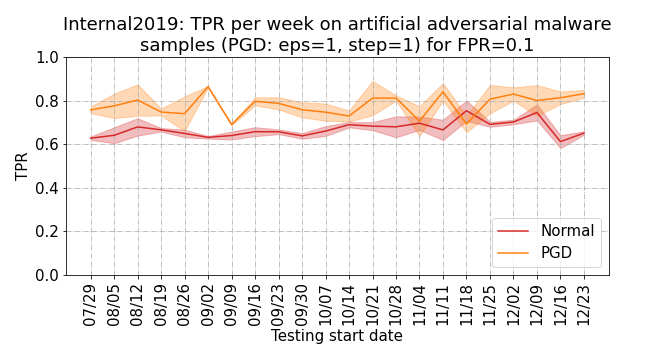}
            \caption{Comparison of TPRs on generated adversarial samples on \textsc{Internal2019} data between normal training and adversarial training using FGSM (left subfigure) and PGD (right subfigure). The results confirm that adversarial training improves robustness of models w.r.t. generated adversarial samples. The shaded regions correspond to the standard error of the mean.}
            \label{fig:robustness}
        \end{figure*}
\begin{figure*}
    \centering
    \begin{subfigure}[b]{0.48\textwidth}
         \centering
          \includegraphics[width=\textwidth]{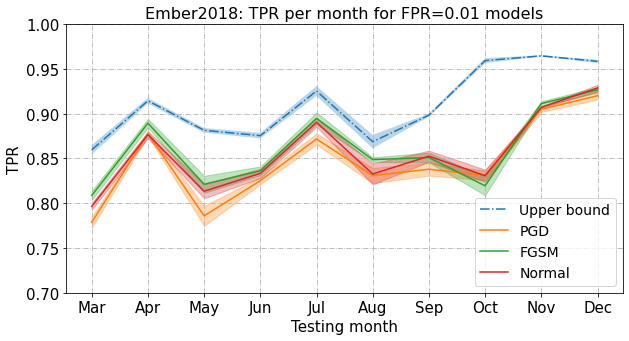}
         \caption{\textsc{Ember2018}}
         \label{fig:ember2018_tpr}
    \end{subfigure}
    \begin{subfigure}[b]{0.48\textwidth}
         \centering
          \includegraphics[width=\textwidth]{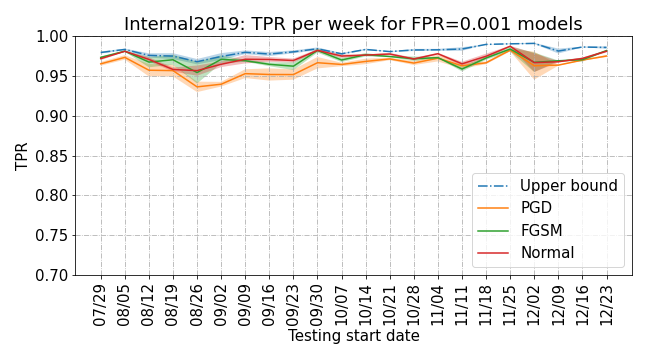}
         \caption{\textsc{Internal2019}}
         \label{fig:internal_tpr}
    \end{subfigure}
    \caption{Comparison of TPRs on testing time-periods for normal and adversarial training on \textsc{Ember2018} (left subfigure) and the \textsc{Internal2019} dataset (right subfigure). The \textsc{UpperBound} corresponds to the (unrealistic) model using data from the future. The shaded regions correspond to the standard error of the mean.}
    \label{fig:TPRs}
\end{figure*}
We now evaluate whether adversarial training methods can yield a model that would improve the TPR in the testing period that follows immediately after the training period. 
To verify that the adversarial training is correct, we first evaluate the TPRs of all models (1 for normal training and 2 for different methods of adversarial training -- using FGSM and PGD) on artificial adversarial samples. 
The evaluation determines whether the models from adversarial training are more robust against these artificially generated adversarial samples.
Results shown in Fig.~\ref{fig:robustness} confirm the expected outcome.
Adversarially trained models have higher TPRs compared to the model with normal training -- for FGSM, this holds across all thresholds, for PGD the difference is noticeable only for larger FPRs. 
While this is most likely due to the small step size for the PGD method, training with larger steps for PGD always yielded significantly worse performance in the subsequent evaluation on future data.

We are now ready to compare the TPRs on the actual testing data.
As discussed in Section~\ref{sec:problem_statement}, we also include the \textsc{UpperBound} model for  reference. 
The \textsc{UpperBound} model demonstrates the possible maximal TPRs one can get when using the best predictor. 
For both datasets, as expected, the \textsc{UpperBound} has the best TPRs (see Fig.~\ref{fig:TPRs}). 
Interestingly, the differences in the TPRs between \textsc{UpperBound} and normal training differ substantially.
For \textsc{Ember2018}, the difference is more apparent. It starts at $\approx5\%$ for the first months and then it is increasing to over $12\%$ advantage for \textsc{UpperBound} in Oct. 
For \textsc{Internal2019}, all models perform considerably better -- note that we have TPRs over $0.95$ for an FPR set to $0.001$ compared to the FPR at $0.01$ and the respective lower TPRs for \textsc{Ember2018}. 
The difference between the normal training and the \textsc{UpperBound} is, however, significantly smaller. 
The main reason lies in the construction of the dataset and the regular retraining. 
Since we constructed \textsc{Internal2019} dataset by subsampling new files incoming to Avast, there is a strong time correlation among the data. 
Moreover, we have a significantly smaller time period (one week compared to one month in \textsc{Ember2018}) during which there does not have to be a significant concept drift among a large fraction of malware samples. 
Therefore, the regularly retrained model performs reasonably well. 
Consequently, there are also only minor differences among the models from normal training compared to the models from adversarial training.
For both datasets, PGD was slightly worse than normal training (the largest difference $\approx 3\%$ was for \textsc{Internal2019} dataset at the beginning of September) while FGSM achieved slightly better TPRs on small FPRs on \textsc{Ember2018} (for March and Apr, FGSM reached higher TPRs by $1.2\%$). However, for the models selected to optimize larger FPRs, FGSM reached slightly lower TPRs (the TPR was lower by $\approx 0.6\%$ on average). 

These experimental results offer two conclusions.
First, in the real-world setting where we assume regular retraining of the models, the negative effects of the concept drift are to some extent contained and there is only a small room for improvement. 
However, in many domains (including the malware detection domain), even a small improvement can have an impact on millions of users, hence it is still valuable to improve the prediction on unseen future data.
Second, adversarial training does not seem to significantly improve the quality of the models.
While for \textsc{Ember2018} the TPRs were marginally higher for models selected to optimize detection for low FPRs, this has not been confirmed on the \textsc{Internal2019} dataset.

To better determine whether adversarial training is really not successful and to decide the same for generative models, we, from now on, focus only on a subset of samples -- namely on malware families that in fact exhibit concept drift. 

~\\
\noindent\textbf{Do Generative Predictions Counteract Concept Drift?}\\
\begin{figure*}[h!]
    \centering
    \includegraphics[width=0.7\textwidth]{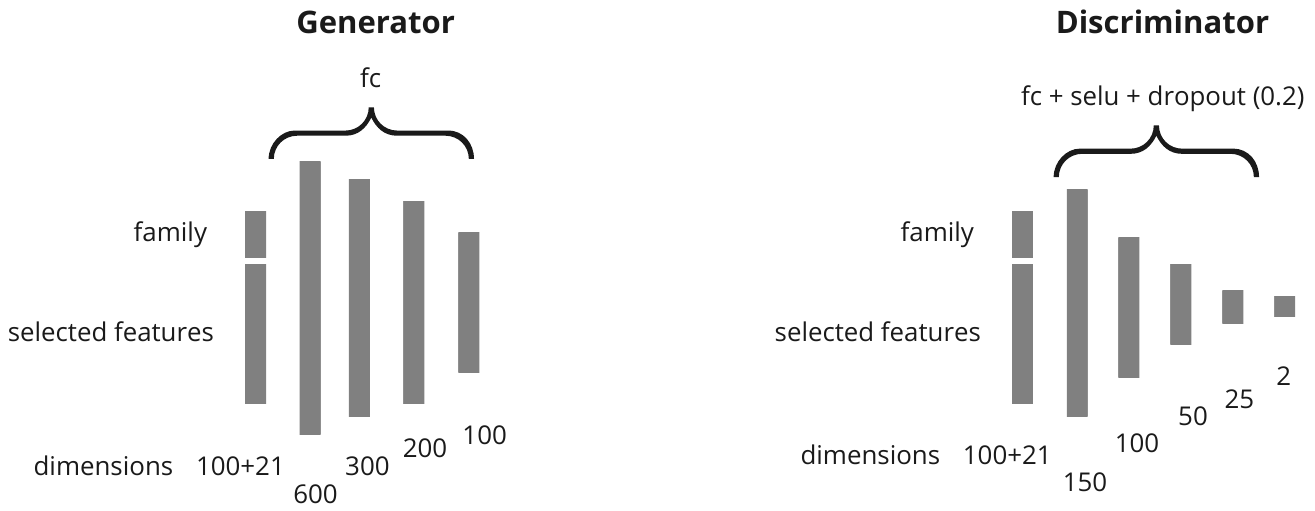}
    \caption{Architectures of neural networks for generator and discriminator. All layers are fully connected (denoted \texttt{fc}), the discriminator uses \texttt{selu} activation function and dropout (with parameter set to 0.2).}
    \label{fig:schemata}
\end{figure*}
~We now turn to the use of the generative models and the described Conditional Cycle GANs (CCyGANs).
Due to computational challenges of training the generative models, we are restricting to a subset of 100 most important features using Lasso feature selection method~\cite{Lasso2014}.
Moreover, given the small difference in performance between \textsc{Normal} training and \textsc{UpperBound} in Fig.\ref{fig:TPRs}, we modify the datasets such that only a subset of malware families will be considered. 
The main reason is that in both datasets there exist malware families that are difficult to predict for the classifier but their data distributions are stable over time (i.e., the prediction difficulty stems from, for example, the closeness of their feature vectors to clean samples).
For such malware families, no prediction method can improve the accuracy on the testing data since the data distributions are the same.
Therefore, we restrict only to those that have the highest variance among data distributions. 

To determine whether data distributions are changing for a malware family, we are using Maximum Mean Discrepancy (MMD) to determine the distance between training and testing data. 
For each dataset and for each time split, we compute MMD for each family. 
Finally, we selected 21 families with the largest sum of all MMD distances. 
The selected families are listed in Table~\ref{tab:selected_families}.

\begin{table}[h]
    \centering
    \begin{tabular}{l  l}
        \toprule 
        \textsc{Ember2018} & \textsc{Internal2019} \\ 
        \midrule
    Vobfus & Adload \\
    Linkury & Trickbot \\
    Spigot & Zeus \\ 
    IStartSurf & Qakbot \\
    Tinba & Laqma \\
    WizzMonetize & Dridex \\
    Sdbot & Bunitu \\
    Ulise &  Emotet \\
    VBClone & Revenge \\
    DownloadHelper & Linkury \\
    Qhost & Phorpiex \\
    FlyStudio & Ursnif \\
    Ainslot & DownloadAssistant \\
    Detnat & FakeAVScan \\
    Soft32Downloader & Sisbot \\
    Kovter & Fearso \\
    Adposhel & Carberp \\
    Qbot & Adposhel\\
    InstallMonster & Kovter \\
    DownloadGuide & Pondfull\\
    Vtflooder & Ulise\\ 
    \bottomrule
    \end{tabular}
    \caption{Selected malware families that exhibit the most significant changes in data distribution.}
    \label{tab:selected_families}
\end{table}

For both datasets, we have used the same architecture of GANs (architectures depicted in Fig.~\ref{fig:schemata}).
The parameters of the neural networks were determined by evaluating multiple settings. 
The input for both generators and discriminators has a dimensionality of 100 + 21 corresponding to 100 reduced features and 21 families in one-hot encoding. Both networks are feed-forward networks with dense layers. The discriminator uses dropout (with rate set to $0.2$) and \texttt{selu} activation function. Finally, for training the classifier, we use a model similar to the discriminator -- the only difference is that the family label is not used as a part of the input (i.e., the input dimensionality is equal to $100$).

We compare 4 different settings: (1) normal training, (2) training with an optimal predictor (\textsc{UpperBound}), (3) adversarial training (using FGSM attacks), and (4) training with samples predicted using generative models.
In each setting, the size of the minibatch was $512$, balanced evenly between clean files and malware files ($256$ samples in each class).
Except for the normal training, in all other cases, the malware part of the minibatch has been split into two parts, where one (size $192$) contained unaltered malware samples from the training set and the second one (size $64$) contained malware samples from differing sources: for the case of (2), malware samples from the future data distributions, for (3), adversarial samples generated by FGSM, and for (4), predictions by CCyGANs. Finally, we have again verified whether the thresholds selected for some particular FPR on the validation data generate comparable FPRs on the testing data for all the compared methods. Similar to the previous setting with the full feature set, the differences among the methods were negligible with one exception -- CCyGANs model had 3-times higher FPR for the last testing week on \textsc{Internal2019} compared to the \textsc{Normal} model.

\subsection{Training CCyGANs}
\begin{figure*}
    \centering
    \begin{subfigure}[b]{0.48\textwidth}
         \centering
         \includegraphics[width=1.0\textwidth]{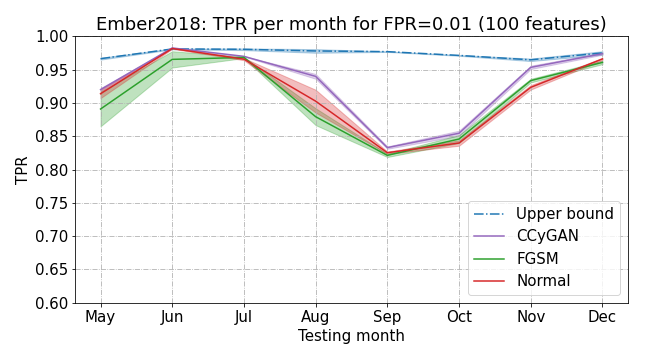}
         \caption{\textsc{Ember2018}}
         \label{fig:ember2018_tpr}
    \end{subfigure}
    \begin{subfigure}[b]{0.48\textwidth}
         \centering
         \includegraphics[width=1.0\textwidth]{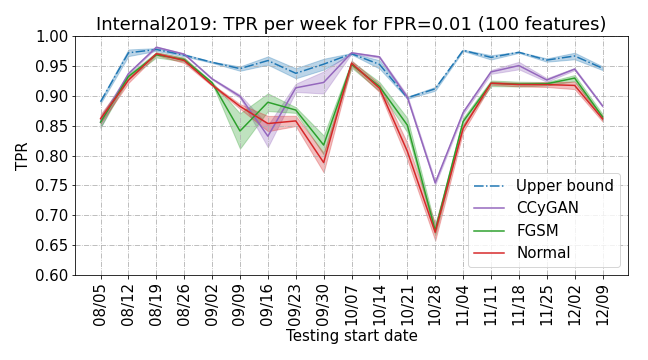}
         \caption{\textsc{Internal2019}}
         \label{fig:internal_tpr}
    \end{subfigure}
    \caption{Comparison of TPRs on testing time-periods for normal,  adversarial training, training with GAN-augmented data (\textsc{CCyGAN}), and (unrealistic) \textsc{UpperBound} classifier on both datasets (left subfigure --  \textsc{Ember2018}; right subfigure -- \textsc{Internal2019}). The shaded regions correspond to the standard error of the mean.}
    \label{fig:TP_GANs}
\end{figure*}
\begin{figure*}
    \centering
    \begin{subfigure}[b]{0.48\textwidth}
         \centering
          \includegraphics[width=1.0\textwidth]{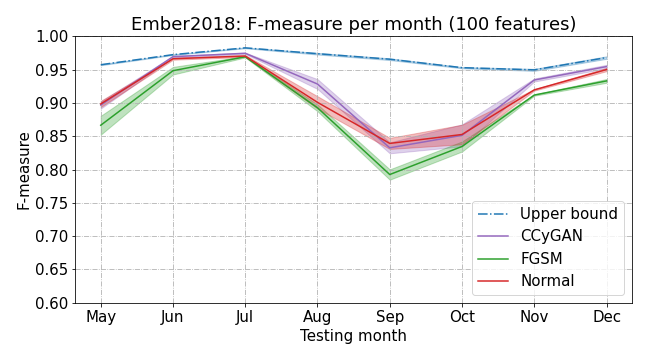}
         \caption{\textsc{Ember2018}}
         \label{fig:ember2018_tpr}
    \end{subfigure}
    \begin{subfigure}[b]{0.48\textwidth}
         \centering
         \includegraphics[width=1.0\textwidth]{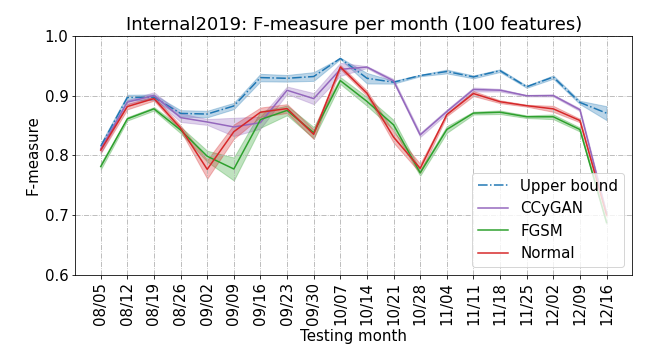}
         \caption{\textsc{Internal2019}}
         \label{fig:internal_tpr}
    \end{subfigure}
    \caption{Comparison of F-measures of all classifiers on both datasets (left subfigure --  \textsc{Ember2018}; right subfigure -- \textsc{Internal2019}). The shaded regions correspond to the standard error of the mean.}
    \label{fig:f1_graphs}
\end{figure*}

As it is known, training GANs can be unstable. 
To improve the convergence, we have used small learning rates ($10^{-4}$ for both generator and discriminator) and alternating two different training periods -- either the parameters of both networks were adapted or only the generator was adapting while the discriminator had been fixed. 
For both datasets, we have used $3500$ training steps with the minibatch of size $512$ and the two training periods were alternating every $50$ training steps (first $50$ training steps, both of the networks are adapting; only the generator is being adapted for the second $50$ training steps, etc.).
We used $L_{cyc}$ loss function and the final reached loss for both generator and discriminator was $\approx\;0.3 - 0.4$. 

In general, determining whether the generators are sufficiently trained is a common problem when using GANs.
However, our usage of GANs is to learn the transformations and then apply these transformations to other data.
Therefore, we have not focused in detail on the evaluation of the quality of trained GANs.
Additional improvements could have been implemented in the training process for GANs, however, the reported method already resulted in a measurable impact on the classifier. 
Hence, any additional improvements in the training process of GANs are out of the scope of this work.

\subsection{CCyGANs Predicions vs. Concept Drift}

Fig.~\ref{fig:TP_GANs} compares the measured TPRs for fixed thresholds on the validation data corresponding to the desired false-positive rate of $0.01$.
The comparison of training the classifier using four mentioned approaches clearly shows that there is a significant space for using predictors for both datasets.
The \textsc{UpperBound} variant with an optimal (but impossible) predictor dominates the other approaches and the differences in the true-positive rates are for some testing periods as high as $15$ \% in September for \textsc{Ember2018} and $24$\% in testing period starting Oct 28 for \textsc{Internal2019}.
On average, including data generated by the \textsc{ub} predictor yields an improvement by $5.9\%$ TPR on \textsc{Ember2018} and by $7\%$ TPR on \textsc{Internal2019}. 

Using data generated by CCyGANs yields promising results. 
On average, the improvements in TPRs are quite small (by $1.3\%$ TPR on \textsc{Ember2018} and by $3.2\%$ TPR on \textsc{Internal2019}, respectively), however, they are consistent -- CCyGAN method is better than normal training for all testing monthts on \textsc{Ember2018} dataset and there are two testing period for \textsc{Internal2019} dataset where CCyGANs have lower average TPR (by $0.5\%$ TPR and by $2\%$) but they are within the intervals of the standard error. 
The gains, on the other hand, are apparent for many testing periods.
CCyGANs were better in August on \textsc{Ember2018} by $3.7\%$ than normal training and by $13\%$ in the testing period starting Sept 30 for \textsc{Internal2019}.

Finally, additional data samples generated by the adversarial method (FGSM) have not proven to act as good predictors. 
For \textsc{Ember2018}, the TPRs are on average worse by $0.6\%$ while there is a small improvement for \textsc{Internal2019} by $0.6\%$.
Especially for \textsc{Ember2018}, adversarial training is sometimes worse and the gains are minimal.
The situation is slightly better for \textsc{Internal2019} dataset, where the most noticable improvements are for testing periods starting in Sept 16 and Sept 23 (by $3.5\%$ and $1.8\%$, respectively).

However, the small gains by FGSM are present only for smaller FPRs.
If we consider models selected for higher FPRs, the gains are no longer present.
On the other hand, the improvement of CCyGANs over the normal training is present regardless of the chosen FPR. 
To provide a more global overview of the performance of the classifiers over multiple FPRs, we have also compared F-measures of the trained classifiers~(see Fig.~\ref{fig:f1_graphs}).
For both datasets, it is clear that training with CCyGAN prediticions improves the F-measure while the F-measure of the adversarially trained classifier is typically worse than the normal training.

\section{Conclusion}
Concept drift is a well-known phenomenon that negatively affects performance of deployed machine-learning models. In many cases, the changes in data distributions happen suddenly and unpredictably and the main mitigation technique is to react to concept drift as quickly as possible by regularly retraining the models/ensembles. 
In other domains, such as in the malware-detection problem, there are fundamental reasons for changes in data distributions.
The goal of the malware authors is to spread their malware and thus we conjecture that the changes in data distributions can be, to some extent, predictable. 
To this end, we introduce a methodology for modeling and discovering changes in malware families -- we train generative adversarial networks to model such changes. 
Next, we use these trained generative networks to \emph{predict} future samples from testing data distributions. 
Our experimental results show that including these artificially generated samples can improve true-positive detection of malware families that exhibit significant concept drift by up to 13\%. 

Our work opens several directions for future work. 
First of all, we have demonstrated the positive impact of using GAN-generated predictions on static features of Windows portable executables. 
However, verifying proposed methodology on the domain of dynamic (behavioral) analysis of executed executables could offer even better results if specific attacking techniques of one malware family are used also in another one. 
Second, while we have demonstrated positive results with our methodology for training and using GANs to predict future samples, a more in-depth analysis of this part could further improve the quality of predicted samples and thus mitigate the negative effects of concept drift even further.
Finally, it would be interesting to move beyond the problem of malware detection to other domains.

\section*{Acknowledgment}
This research was partially supported by the OP VVV MEYS funded project CZ.02.1.01/0.0/0.0/16 019/0000765 ``Research Center for Informatics''.

\ifCLASSOPTIONcaptionsoff
  \newpage
\fi

\bibliographystyle{IEEEtran}

\end{document}